\RequirePackage{fix-cm}

\documentclass[onecolumn]{svjour3}          
\usepackage{amsmath,amssymb}
\smartqed  
\usepackage{url}
\usepackage[%
  colorlinks=true,
  urlcolor=blue,
  linkcolor=blue,
  citecolor=blue
]{hyperref}
\hypersetup{linkcolor=red, citecolor=red, colorlinks=true}

\usepackage{mathtools,amsmath}

\usepackage{ulem}

\begin{document}

\title{Nonequilibrium phase transitions in a racism-spreading model with interaction-driven dynamics}




\author{Nuno Crokidakis  \and Lucas Sigaud}


\institute{
           Nuno Crokidakis \at
           Instituto de F\'isica, Universidade Federal Fluminense, Niter\'oi/RJ, Brazil \\
           \email{nunocrokidakis@id.uff.br} \\
           \url{https://scholar.google.com/citations?user=Oy-opMUAAAAJ&hl=pt-BR&oi=ao}
           \and
           Lucas Sigaud \at
           Instituto de F\'isica, Universidade Federal Fluminense, Niter\'oi/RJ, Brazil \\
           \url{https://scholar.google.com/citations?user=3NFG9RIAAAAJ&hl=pt-BR}
}

\date{Received: date / Accepted: date}

\maketitle

\begin{abstract}
Racism remains a persistent societal issue, increasingly amplified by the structure and dynamics of online social networks. In this work, we propose a three-state compartmental model to study the spreading and suppression of racist content, drawing from epidemic-like dynamics and interaction-driven transitions. We analyze the model on fully-connected (homogeneous mixing) networks using a set of coupled differential equations, and on Barabási-Albert (BA) scale-free and Watts-Strogatz (WS) small-world networks through agent-based simulations. The system exhibits three distinct stationary regimes: two racism-free absorbing states and one active phase with persistent racist content. We identify and characterize the phase transitions between these regimes, discuss the role of network topology, and highlight the emergence of absorbing states. Our findings illustrate how statistical physics tools can help uncover the macroscopic consequences of microscopic social interactions in digital environments.

\end{abstract}

\section{Introduction}

\quad The propagation of racist messages through social networks has become an increasingly pressing issue, prompting researchers in many different fields, and with different approaches, to develop mathematical models to understand and mitigate this phenomenon \cite{balak,kamran,vera,bliuc}. A key reference in this domain is the work on the Susceptible-Exposed-Infected-Denier (SEID) model, which describes the spread of racism online by treating racism as a social disease and classifying individuals into susceptible, exposed, infected, and denying categories. This approach highlights the stages through which users might encounter and potentially perpetuate racist content, offering a framework to study interventions that could disrupt this process \cite{mamo1}.

Statistical physics has proven to be a powerful framework for modeling social phenomena, particularly those involving large-scale emergent behavior from simple local interactions \cite{rmp,caos}. Topics such as opinion dynamics, cultural diffusion, and collective decision-making have all been successfully analyzed using tools from nonequilibrium statistical mechanics and complex network theory. In this context, the spread of racist ideologies can be understood as a contagion-like process driven by interpersonal influence and shaped by network topology. This motivates the development of models that capture both the stochastic nature of individual interactions and the global patterns that arise from them \cite{sooknanan}.

A further noteworthy contribution is a work dealing with the interplay between community resilience and the spread of racism. This study incorporates resilience factors into mathematical models, providing insights into how community characteristics can influence the dynamics of hate speech dissemination. By identifying conditions under which communities are more resistant to racist ideologies, this work lays the groundwork for targeted prevention strategies \cite{mamo2}.

The application of agent-based models to simulate hate speech propagation is also a significant development in this field. The study demonstrates how individual-level behaviors aggregate to influence the macro-level spread of hate messages \cite{mamo3}. This bottom-up modeling approach is particularly useful for examining the role of influencers, echo chambers, and counter-speech in shaping the trajectory of racist content within digital platforms.

In parallel, the use of natural language processing and network analysis to model the dynamics of racist online communities offers a complementary perspective. The pre-print by Arviv \textit{et. al.} reconstructs the networks of these communities and analyzes their growth and inter-connectivity. This line of inquiry underscores the importance of structural features in amplifying or mitigating the spread of harmful ideologies \cite{arviv}.

The issue of racism is not confined to social media but is also prevalent in sports, as demonstrated by incidents involving prominent football players. A notable and recent example is the case of Vinícius Júnior, a Real Madrid Football Club forward, who has faced multiple instances of racial abuse during matches. In particular, on May 21, 2023, during a game against Valencia Football Club, he was subjected to racist chants from the crowd, sparking widespread condemnation. This incident led to legal consequences, with three Valencia fans being sentenced to eight months in prison for hate crimes \cite{Reuters2024}.

In response to ongoing racial abuse, Vinícius Júnior and many other players have been vocal about the lack of robust institutional responses to such incidents. His case has highlighted systemic issues within football organizations, pushing for stronger actions against racism in sports. Recent reports suggest continued efforts to address these challenges, with legal proceedings and public awareness campaigns gaining momentum \cite{Reuters2024,ESPN2023}. In many cases, passionate behavior towards something such as a sports team may trigger what has been called "frozen prejudices" \cite{galam_trump1,galam_trump2,galam_entropy_2025}, which can manifest in these environments as a byproduct of herd-like fan behavior and then propagate via digital media to broader audiences.

Beyond academic studies, recent news articles provide real-world context to these theoretical models. For instance, reports on coordinated racist text messages sent across the United States illuminate the mechanisms by which racism can manifest in digital communication. These incidents underscore the relevance of understanding how messages travel through networks, as well as the potential for rapid and widespread dissemination \cite{news1,news2,news3,news4,news5}. 

A recent study adds another dimension to this body of work by exploring the intersection of racism and corruption \cite{kotola}. Using a deterministic compartmental model, the authors analyze how these two social issues can coexist and reinforce each other within communities. This approach highlights the feedback loops and systemic factors that sustain these problems, emphasizing the need for integrated strategies that address both racism and corruption simultaneously. By providing a mathematical framework to study these interactions, the research offers a novel perspective on tackling deeply rooted societal challenges.

In contrast to previous models of racism propagation based on epidemic-like frameworks such as the SEID model \cite{mamo1} and its variants \cite{mamo2,mamo3}, our approach introduces several key innovations. First, we emphasize interaction-driven transitions, reflecting the central role of social contact in online platforms, rather than relying on spontaneous changes of state. Second, we analyze the system's nonequilibrium behavior, including the identification of absorbing states and continuous phase transitions, which were not addressed in those earlier studies. Third, we extend the model to heterogeneous network topologies (specifically, Barab\'asi-Albert networks), allowing us to capture the influence of real-world connectivity patterns on the spread and control of racist ideologies. These contributions differentiate our model both conceptually and methodologically from prior work. Furthermore, we aim to propose a model with the least number of free parameters, in order to evaluate the main mechanisms leading to its final states.

In contrast to previous epidemic-based models of racism propagation, such as the SEID model and its extensions, our approach introduces key conceptual and dynamical innovations. By relying exclusively on interaction-driven transitions, we emphasize the role of social contact in state changes. Our framework also reveals absorbing states and nonequilibrium phase transitions, which are not typically addressed in previous models. Additionally, we analyze both homogeneous and heterogeneous network topologies, exploring how structural connectivity affects the resilience or fragility of racism-free states. These contributions situate our work within the broader tradition of statistical physics applied to complex social phenomena \cite{rmp,caos}.

To accomplish this, we propose a three-state compartmental model to analyze the spread of racism on social networks. We begin by constructing a model on a fully-connected graph, where transitions between states are governed by defined probabilities, leading to a set of rate equations. To better capture the complex topology of real-world social networks, we extend our analysis by implementing the model both on Barabási-Albert (BA) and Watts-Strogatz (WS) small-world networks. Our findings indicate the potential coexistence of two or three subpopulations, as well as the presence of an absorbing state. Furthermore, we explore conditions under which the spread of racism can be effectively controlled, offering insights into possible interventions.


\section{Model}

\qquad Before presenting the equations and dynamics, we highlight the main conceptual features that distinguish our model from previous approaches such as SEID-type frameworks and SIS models. First, transitions between compartments are not spontaneous but are induced solely through interactions, better reflecting the dynamics of digital platforms. Second, we introduce a Denier (D) state with absorbing character, capturing individuals who permanently reject racist content and can no longer be re-infected. This mechanism leads to nontrivial dynamical behavior, including absorbing configurations and continuous phase transitions. Finally, by avoiding additional intermediate compartments and free parameters, we aim to retain a minimal model structure that still captures rich collective phenomena and, as a consequence of employing less states, emphasizes the main driving aspects of the system.

\quad In this work, we consider a three-compartment diffusion model with a population of $N$ individuals (henceforth called agents) on a given network of contacts. Each agent must belong to one of the three possible compartments, namely Susceptibles ($S$), Infected ($I$) and Deniers ($D$). For the purposes of this work, we restrict our analysis to social media contact, under the assumption that in-person interactions are either simultaneous with, or significantly less far-reaching than, those occurring on social media. Agents in compartment $S$ are individuals who have never had any contact whatsoever with racism, and therefore are susceptible to be influenced by both racist and nonracist groups. Agents in groups $I$ and $D$ both have already been exposed to racism (via interactions described below), but with different outcomes: $I$ individuals believe in the racist propaganda they received, and consequently act as racism-spreading agents, while the population in $D$ denies racism and operates in ways to try to quench its spreading.

Transitions may be mediated by interactions from agents from one compartment to another, as well as by births and deaths. Regarding the latter, new individuals (either recently born or coming-of-age) can only enter the system via compartment $S$, with rate $\Lambda$, as we assume that all new individuals had not a chance yet to be exposed to racism; meanwhile, individuals can die in any of the three groups. For simplicity, we assume that the death rate of all compartments are the same: $\mu$.

\begin{figure}[t]
\begin{center}
\vspace{6mm}
\includegraphics[width=0.55\textwidth,angle=0]{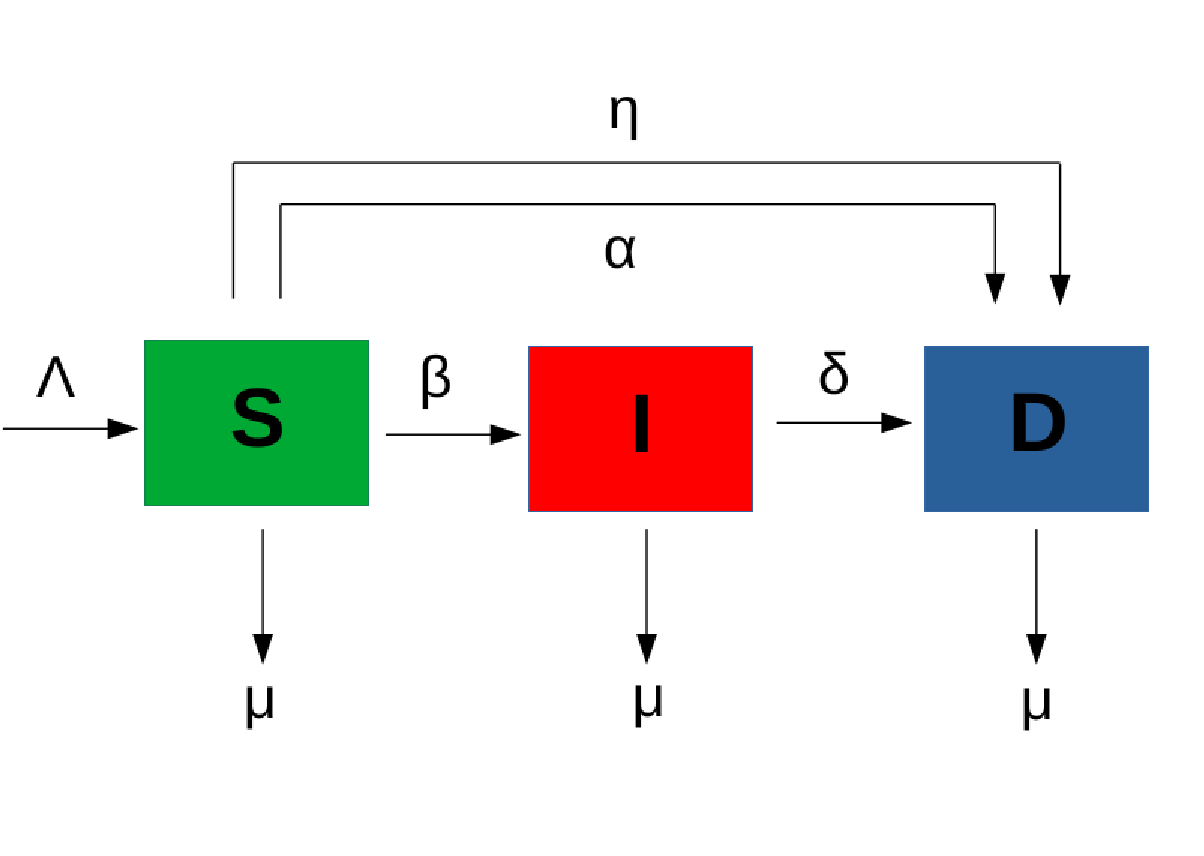}
\end{center}
\caption{Illustration of the model's compartments: Susceptible (S), Infected (I) and Deniers (D).}
\label{fig1}
\end{figure}

Agents in the $S$ compartment can migrate to the other two compartments when exposed to racism. We consider both cases where contact with a certain type of agent may sway an $S$ individual towards the agent's point of view - in other words, interaction between an $S$ and an $I$ individual may turn the susceptible one into an infected agent (with probability $\beta$), while if an $S$ individual interacts with a $D$ agent, they may become a denier (with probability $\eta$). Infected individuals, if they are exposed to deniers, can also be convinced in our model to change their points of view and become against racism, with probability $\delta$. We also consider the possibility that a susceptible individual, even if exposed to racist content via infected agents, denies racism due to their own accord and move directly to the $D$ compartment, with probability $\alpha$. These interactions are mapped on the reaction equations below for each group and the entire model is pictorially represented in Fig. \ref{fig1}. Table \ref{tab:params} provides a comprehensive summary of all parameters used in the model.

\begin{table}[h]
\centering
\caption{Model parameters and their interpretation.}
\begin{tabular}{|c|l|}
\hline
\textbf{Parameter} & \textbf{Interpretation} \\
\hline
\( \beta \) & Probability that a Susceptible becomes Infected after interacting with an Infected individual. \\
\( \alpha \) & Probability that a Susceptible becomes a Denier after interacting with an Infected individual. \\
\( \eta \) & Probability that a Susceptible becomes a Denier after interacting with a Denier. \\
\( \delta \) & Probability that an Infected becomes a Denier after interacting with a Denier. \\
\( \mu \) & death rate \\
\( \Lambda \) & birth rate \\
\hline
\end{tabular}
\label{tab:params}
\end{table}

\begin{eqnarray} \label{eq1}
    S + I & \stackrel{\beta}{\rightarrow} & I + I, \\ \label{eq2}
    S + I &\stackrel{\alpha}{\rightarrow} & D + I, \\ \label{eq3}
    S + D & \stackrel{\eta}{\rightarrow} & D + D, \\ \label{eq4}
    I + D & \stackrel{\delta}{\rightarrow} & D + D.
\end{eqnarray}

The model is structurally similar to many SIR-type epidemic models that have been extensively used over the past decades - particularly during the Covid-19 pandemic - to describe the propagation of contact-based diseases \cite{anderson,nuno1,nuno2}. However, the underlying dynamics in our case is markedly different, due to the specific interaction rules among groups. In particular, a susceptible individual may transition to the denier state through two distinct pathways: either by interacting with an infected or a denier agent. We emphasize the role of all key interaction mechanisms while keeping the number of free parameters minimal, in order to maintain the model as simple and realistic as possible.

It is worth noting that the present model can also be interpreted in the broader context of competitive spreading processes on networks. Indeed, our model exhibits structural analogies with competitive spreading processes on complex networks, as reviewed in Pastor-Satorras et al. \cite{rmp_epidemic}. In particular, the coexistence and transitions among the susceptible, infected, and resistant subpopulations resemble dynamics observed in multi-strain epidemic models or in awareness-disease competition frameworks. The resistant class functions as an inhibitory mechanism analogous to immunized or informed agents, introducing nontrivial feedback that suppresses the spread of the primary contagion - in this case, racist ideology. Such antagonistic interactions give rise to rich dynamical behavior, including multiple stable fixed points and critical thresholds, features that are well known in the context of competing contagions. Our formulation captures these mechanisms through nonlinear interaction terms and transition probabilities, leading to bifurcation phenomena and phase transitions that mirror those found in epidemic models with interacting processes.

In order to explore these features, in the next section the dynamics of the model described here is explored in three different networks: (i) a fully-connected network, which corresponds to a mean-field statistical model where all encounters are equally probable and the differences between the interaction probability parameters are its only determinant features; (ii) a Barab\'asi-Albert network, where the interactions depend on the separation between the individual and the other connected nodes in the network and the degree distribution follows a power law \cite{barabasi1,barabasi2}; and (iii) a Watts-Strogatz network, where individuals are initially connected in a regular lattice and some links are randomly rewired with probability $p$, generating a small-world topology characterized by high clustering and short average path lengths \cite{watts1998collective}.


\section{Results}

\subsection{Fully-connected network}

\quad In this subsection we consider the model defined in section 2 in a fully-connected population. Considering for simplicity $\Lambda=\mu\,N$, i.e. the overall number of individuals does not change over time, and taking into account the subpopulation fractions $s=S/N, i=I/N$ and $d=D/N$, one can translate the rules discussed previously in mean-field rate equations, given by
\begin{eqnarray}\label{eq5}
\frac{d}{dt}\,s & = & \mu -(\beta+\alpha)\,s\,i - \eta\,s\,d - \mu\,s \\ \label{eq6}
\frac{d}{dt}\,i & = & \beta\,s\,i - \delta\,i\,d - \mu\,i \\ \label{eq7}
\frac{d}{dt}\,d & = & \delta\,i\,d + \eta\,s\,d + \alpha\,s\,i  - \mu\,d 
\end{eqnarray}
\noindent
Observe that Eqs. (\ref{eq5}) to (\ref{eq7}) obey the condition $s+i+d=1$. These mean-field equations correspond to the fully-connected (homogeneous mixing) limit of the model.

In this framework, susceptibility is represented by the population in the $S$ compartment and their potential to change state due to interactions. Exposure corresponds to encounters between susceptible individuals and agents in the $I$ (infected) or $D$ (denier) states, which may induce transitions. This is quantitatively described in the model by interaction terms proportional to $\beta\,s\,i$ and $\alpha\,s\,d$, respectively. Influence is encoded in the values of the transition parameters $\beta$ and $\alpha$, which control the strength of transmission from infected and denier individuals, respectively, to susceptible ones. These mechanisms jointly govern the dynamics through the nonlinear interaction terms in Eqs. (\ref{eq5})–(\ref{eq7}).

\begin{figure}[t]
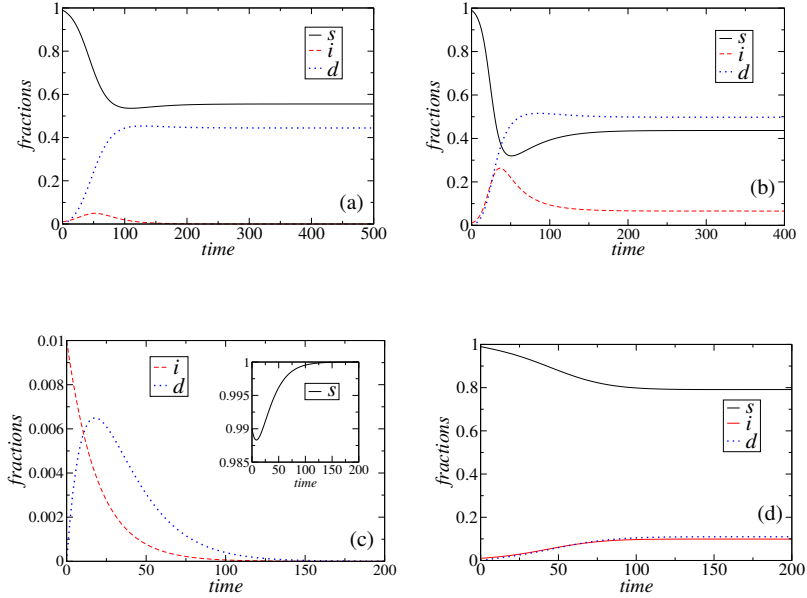

\begin{center}
\vspace{6mm}
\includegraphics[width=0.4\textwidth,angle=0]{figure2a.eps}
\hspace{0.3cm}
\includegraphics[width=0.4\textwidth,angle=0]{figure2b.eps}
\\
\vspace{1.0cm}
\includegraphics[width=0.42\textwidth,angle=0]{figure2c.eps}
\hspace{0.3cm}
\includegraphics[width=0.4\textwidth,angle=0]{figure2d.eps}
\end{center}
\caption{Time evolution of the agents' fractions in the three compartments $s, i$ and $d$ for the mean-field formulation of the model, obtained from the numerical integration of Eqs. (\ref{eq5}) - (\ref{eq7}). The fixed parameters are $\alpha=0.20, \delta=0.15$ and $\eta=0.18$, and we varied $\beta$ and $\mu$: (a) $\mu=0.10, \beta=0.20$, (b) $\mu=0.10, \beta=0.40$, (c) $\mu=0.30, \beta=0.20$ and (d) $\mu=0.30, \beta=0.40$.}
\label{fig2}
\end{figure}

The evolution of fractions $s(t), i(t)$ and $d(t)$ over time can be obtained from the numerical integration of Eqs. (\ref{eq5}) - (\ref{eq7}). In Fig. \ref{fig2} we exhibit results for the initial condition $s(0)=0.99, i(0)=0.01, d(0)=0.00$ and fixed parameters $\alpha=0.20, \delta=0.15$ and $\eta=0.18$ \footnote{Since empirical data for these parameters are limited or difficult to obtain, we explore the model’s dynamics over a representative range of values to uncover the qualitative regimes and phase transitions.}. In this case, we varied the contagion rate $\beta$ and the birth/death rate $\mu$. It is crucial to notice the importance of parameters $\beta$ and $\mu$ to the overall behavior of the system - for larger birth/death rates (Fig. \ref{fig2} (c) and (d), for $\beta$ values 0.2 and 0.4, respectively), it is clear that the susceptible individuals remain that most common ones in the system, reaching an absorbing state for the lower contagion rate (0.2) - beware the scale change for Fig. \ref{fig2} (c) -  and reaching an equilibrium at 80\% population for a higher one (0.4). For the latter case, racism spreaders and deniers remain both at $\sim$10\%. This is not unexpected: a high $\mu$ means a higher ``refresh rate'' of the system, since all new individuals start at the $S$ compartment, but are proportionally removed from all three. It is worthwhile to point out that this ``refresh rate'' is also the only means by which the $S$ population can grow - by construction, if an agent has contact with racism in any manner, it becomes a racism spreader or denier, but never again a susceptible; therefore, no interaction can lead to an individual becoming part of the $S$ subgroup again. On the other hand, lower $\mu$ leads to more significant fractions of the other groups, which can be more sensitive to variations of the $\beta$ contagion parameter, which are further explored using the steady states of the model.

Considering the steady states of the model, Eq. (\ref{eq6}) gives us two solutions (see the Appendix for the calculations). One of the solutions is $i=0$, which represents a racism-free stationary state. Taking this racism-free stationary solution into account, we can show that there are two distinct collective steady states associated with it (see the Appendix). The first solution is $s_1=1,i_1=d_1=0$, which represents an absorbing state where only susceptible agents can be found in the population. The second solution associated with $i=0$ is given by $s_2=\mu/\eta, i_2=0$ and $d_2=1-\mu/\eta$. This last solution represents a stationary state where the populations of susceptible and denier agents coexist, despite the extinction of the racism spreaders $i$. 

Following the stability analysis presented in the Appendix A, we can see that there are two distinct nonequilibrium critical points in the model, namely
\begin{eqnarray} \label{betac1}
\beta_c^{(1)} & = & \mu \\   \label{betac2}  
\beta_c^{(2)} & = & (\delta+\mu)\,\frac{\eta}{\mu} - \delta
\end{eqnarray}

\noindent
The absorbing state solution will be valid for $\mu>\eta$, since $\beta<\beta_c^{(1)}$. On the other hand, the other solution is valid for $\mu<\eta$, since $\beta<\beta_c^{(2)}$. Equating the expressions for the two critical points, they coalesce at $\mu=\eta$. This value separates both solutions with $i=0$.

In addition to the above two solutions regarding $i=0$, there is also a solution with $i\neq 0$, valid for $\beta>\beta_c^{(1)}$ and $\beta>\beta_c^{(2)}$ (see the Appendix). This solution is given by 
\begin{eqnarray} \label{eq8_1}
s_3 & = & \frac{\delta}{\beta}\left(\frac{\beta-\mu}{\beta+\delta} + \frac{\mu}{\delta}\right) + \beta\,i_{3}^{*} \\ \label{eq8_2}
i_3 & = & i_{3}^{*} \\ \label{eq8_3}
d_3 & = & \frac{\beta-\mu-\beta\,i_{3}^{*}}{\beta+\delta} 
\end{eqnarray}
\noindent
where $i_{3}^{*}\neq 0$. This solution represents an active phase where the three densities $s, i$ and $d$ are present in the steady states of the model. In particular, it is the only phase where there is an endemic survival of racism, i.e., a certain fraction of racist agents $i\neq 0$ are present in the steady states of the system. Thus, we expect to observe three distinct nonequilibrium phase transitions. One of the transitions occurs between the absorbing and the active phases, a second transition between the phase where $s$ and $d$ coexist and the active phase, and the last one between the absorbing phase and the phase where $i=0$ and $s$ and $d$ coexist.

\begin{figure}[t]
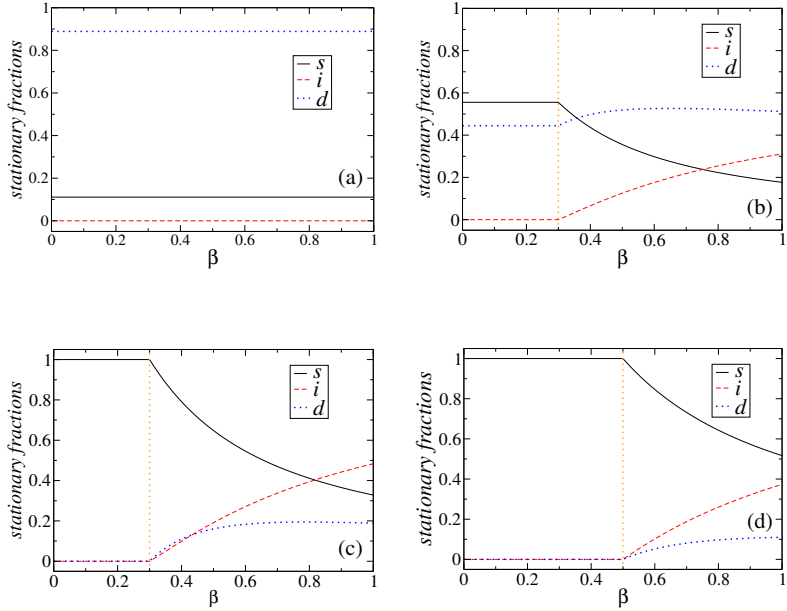

\begin{center}
\vspace{6mm}
\includegraphics[width=0.4\textwidth,angle=0]{figure3a.eps}
\hspace{0.3cm}
\includegraphics[width=0.4\textwidth,angle=0]{figure3b.eps}
\\
\vspace{1.0cm}
\includegraphics[width=0.4\textwidth,angle=0]{figure3c.eps}
\hspace{0.3cm}
\includegraphics[width=0.4\textwidth,angle=0]{figure3d.eps}
\end{center}
\caption{Stationary fractions $s, i$ and $d$ as functions of the contagion probability $\beta$ for the mean-field formulation of the model, obtained from the numerical integration of Eqs. (\ref{eq5}) - (\ref{eq7}). The fixed parameters are $\alpha=0.20, \delta=0.15$ and $\eta=0.18$, and we varied $\mu$: (a) $\mu=0.02$, (b) $\mu=0.10$, (c) $\mu=0.30$ and (d) $\mu=0.50$. The vertical (orange) dotted lines in panels (b) - (d) indicate analytically predicted critical values $\beta_{c}^{(2)}$ (panel (b)) and $\beta_{c}^{(1)}$ (panels (c) and (d)) associated with phase transitions (see the main text).}
\label{fig3}
\end{figure}

To better visualize this situation, we plot in Fig. \ref{fig3} the stationary fractions $s, i$ and $d$ as functions of the contagion probability $\beta$. The fixed parameters are $\alpha=0.20, \delta=0.15$ and $\eta=0.18$, and we varied the birth/death rate $\mu$. One can clearly see the onset of the steady-state $i$ population for $\beta$ higher than a critical value. On the other hand, it must be pointed out that the overall picture varies considerably with the parameter $\mu$. With the increase of the $\mu$ parameter, the deniers population (predominant throughout all values of $\beta$ for $\mu = 0.02$) tend to decrease, while both the susceptibles and propagators increase, at least for a while. For high values of $\mu$ (e.g. $\mu = 0.50$ at the panel (d) of Fig. \ref{fig3}), though, due to the increase of the ``refresh rate'' discussed above, the susceptibles become predominant, regardless of the values of $\beta$.

\begin{figure}[t]
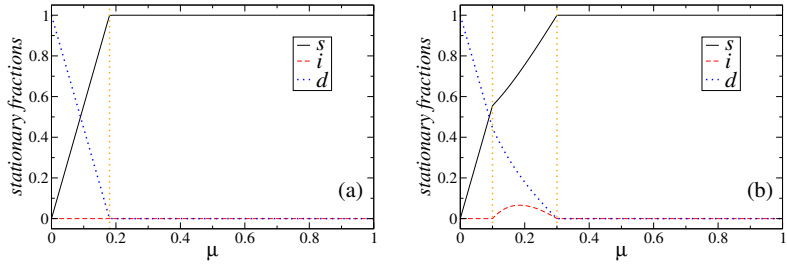

\begin{center}
\vspace{6mm}
\includegraphics[width=0.4\textwidth,angle=0]{figure4a.eps}
\hspace{0.3cm}
\includegraphics[width=0.4\textwidth,angle=0]{figure4b.eps}
\end{center}
\caption{Stationary fractions $s, i$ and $d$ as functions of the birth/death probability $\mu$ for the mean-field formulation of the model, obtained from the numerical integration of Eqs. (\ref{eq5}) - (\ref{eq7}). The fixed parameters are $\alpha=0.20, \delta=0.15$ and $\eta=0.18$, for two distinct values of $\beta$: (a) $\beta=0.10$ and (b) $\beta=0.30$. The vertical (orange) dotted lines indicate analytically predicted critical values associated with phase transitions (see the main text).}
\label{fig4}
\end{figure}

\begin{figure}[t]
\begin{center}
\vspace{8mm}
\includegraphics[width=0.65\textwidth,angle=0]{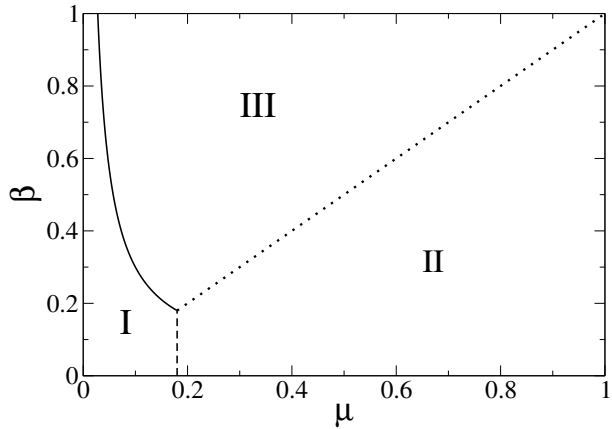}
\end{center}
\caption{Phase diagram of the model on the plane $\beta$ versus $\mu$. The fixed parameters are $\alpha=0.20, \delta=0.15$ and $\eta=0.18$. The diagram show the three macroscopic collective states, represented by phases \textbf{I, II} and \textbf{III}. The lines denote the critical points $\beta_c^{(1)}$ (dotted line) and $\beta_c^{(2)}$ (full line) given by Eqs. (\ref{betac1}) and (\ref{betac2}), respectively, as well the special value of $\mu=\eta$ where both critical points are equal (dashed line). This last line separates the two racism-free stationary solutions.}
\label{fig5}
\end{figure}

This becomes even more evident when we analyze the stationary fractions as functions of the birth/death rate $\mu$. In Fig. \ref{fig4} we exhibit the fractions $s, i$ and $d$ for fixed parameters $\alpha=0.20, \delta=0.15$ and $\eta=0.18$, and two distinct values of $\beta$, namely $\beta=0.10$ (panel (a)) and $\beta=0.30$ (panel (b)). The $s$ compartment becomes dominant quickly with the increase of $\mu$, and once again we can notice that there is a resiliency of the $i$ group only for configurations with $\beta > \beta_c^{(1)},\beta_c^{(2)}$ and $\mu$ between two critical values. In other words, for small values of $\beta$, as we increase the value $\mu$ we transition from one phase where $(s,i,d) = (\mu/\eta, 0, 1-\mu/\eta)$ to an absorbing one where $(s,i,d)=(1,0,0)$, at the critical point $\mu_c=\eta$ (see the vertical dotted line in Fig. \ref{fig4} (a)). Looking at the phase diagram shown in Fig. \ref{fig5}, this is represented at the low portion of the diagram by the crossing of the vertical dashed line separating areas $I$ and $II$. Meanwhile, for higher values of $\beta$ (for the parameters of Fig. $\ref{fig5}$, as low as 0.2), three distinct phases can be observed. For $\mu<\mu_c$, we once again obtain phase $I$, or $(s,i,d) = (\mu/\eta, 0, 1-\mu/\eta)$. This $\mu_c$ can be obtained from the expression for $\beta_c^{(2)}$ (see the Appendix), resulting in $\mu_c=(\delta\,\eta)/(\beta+\delta-\eta)$, which for the parameters of Fig. \ref{fig4} (b) gives $\mu_c=0.1$ (see one of the vertical dotted lines in Fig. \ref{fig4} (b)). Since $\beta_c^{(1)}=\mu$, the absorbing phase ($II$) is reached for $\mu>\beta$ ($0.3$ for Fig. \ref{fig4} (b), see the other vertical dotted line). The active phase ($III$ in the diagram of Fig. \ref{fig5}), where $s,i,d \neq 0$, therefore exists for $\mu_c < \mu < \beta_c^{(1)}$.

Summarizing the previous discussion, we plot in Fig. \ref{fig5} the phase diagram of the model on the plane $\beta$ versus $\mu$. The fixed parameters are $\alpha=0.20, \delta=0.15$ and $\eta=0.18$.  The curve for small values of $\mu$ is the critical point $\beta_c^{(2)}$ (full line), wheres the straight (dotted) line for $\mu>\eta=0.18$ represents the critical point $\beta_c^{(1)}=\mu$. Phase \textbf{I} represents the collective state where the subpopulations $s$ and $d$ coexist and the racism spreaders $i$ disappear of the population. Phase \textbf{II} indicates the absorbing states of the model ($s=1$ and $i=d=0$), and the area represented by \textbf{III} denotes the active phase, where the three compartments coexist in the stationary states. The vertical (dashed) line represents the special value $\mu=\eta=0.18$, that separates both racism-free solutions ($i=0$).

We remark that these transitions are not merely empirical observations from numerical integration. Rather, they are analytically grounded: the epidemic thresholds $\beta_c^{(1)}=\mu$ and $\beta_c^{(2)}=(\delta+\mu)\,\frac{\eta}{\mu} - \delta$ are obtained from linear stability analysis around the fixed points, as shown in Appendix A. The transitions between the absorbing, coexistence, and active phases are all continuous, with no evidence of hysteresis or bistability in the mean-field equations. While we did not compute critical exponents, the continuous nature of the transitions and the structure of the model are consistent with the directed percolation universality class,  as commonly found in spreading models with absorbing states and continuous phase transitions \cite{dickman}. While we did not compute critical exponents, the continuous nature of the transitions and the structure of the model are consistent with the directed percolation (DP) universality class, which is commonly found in systems with absorbing states and no conservation laws or additional symmetries \cite{dickman}. In particular, our model in the fully-connected case exhibits a unique absorbing state, stochastic dynamics and no quenched disorder, placing it within the class of models that are conjectured to belong to DP according to general criteria \cite{Hinrichsen,Odor}. This connection provides a useful conceptual framework, even in the absence of finite-size scaling analysis.

\subsection{Barab\'asi-Albert (BA) network}

\qquad Recent developments in network science have provided powerful tools to model social contagion processes, including the spread of ideologies, behaviors and misinformation across heterogeneous structures. Several studies have leveraged complex network models to investigate how topological features influence the propagation of information and collective dynamics \cite{Dong,Liu,Qing}. These approaches underscore the importance of considering nontrivial connectivity patterns when analyzing social phenomena in digital environments.

\quad Thus, in this section we will consider a more realistic situation, where the network of social contacts is represented by a BA network. We generated BA networks with $N$ nodes in the standard way, i.e., the social network follows the BA model, where nodes (representing individuals) are connected based on a preferential attachment mechanism, leading to a scale-free topology \cite{barabasi1}. The BA network was generated starting from an initial configuration of $m_o=2$ connected nodes. At each time step, a new node is added to the network and connected to one existing node ($m=1$) with probability proportional to the degree of the existing nodes (i.e., preferential attachment). Each independent trial uses the same BA network instance. The two seed nodes in the initial configuration are connected by a single edge.

Each agent can be in one of the three possible states, namely $S, I$ or $D$. The model presented in section 2 is then simulated according to the following steps:

\begin{itemize}

\item Initial Conditions: Initially, we randomly assign a fraction of the total nodes $N$ to start in the $I$ state (individuals actively promoting racist views). The remaining nodes begin in the susceptible ($S$) state, depending on predefined fractions;

\item Interaction Dynamics: At each time step, a given individual interacts  with one of his/her nearest neighbors, chosen at random, and the state of the node can change according to predefined transition probabilities, as we pointed in Eqs. (\ref{eq1}) - (\ref{eq4});

\item Spreading Mechanisms:

\begin{itemize}

\item If a susceptible ($S$) agent is in contact with an infected ($I$) neighbor, he/she can transition to the $I$ state with probability $\beta$ or to the $D$ state with probability $\alpha$;

\item If a susceptible ($S$) agent is in contact with an denier ($D$) neighbor, he/she can transition to the $D$ state with probability $\eta$;

\item In an infected ($I$) agent is in contact with an denier ($D$) neighbor, he/she can transition to the $D$ state with probability $\delta$;

\end{itemize}

\item Every node is visited once per time step, and an attempt is made to update its state based on predefined transition rules. One full time step in the model corresponds to $N$ such attempts, ensuring that, on average, each node has a chance to update its state once per step.

\end{itemize}

For the simulations performed here, we considered BA networks with a size of $N=5000$ nodes. At $t=0$, $2\%$ of all agents are in the $I$ state, randomically distributed in the network, while the remaining start at the $S$ state. Although the BA network is naturally more realistic than the fully-connected one, the temporal evolution of the BA network is very similar to the ones shown in Fig. \ref{fig2}, and are therefore not shown here. 

\begin{figure}[t]
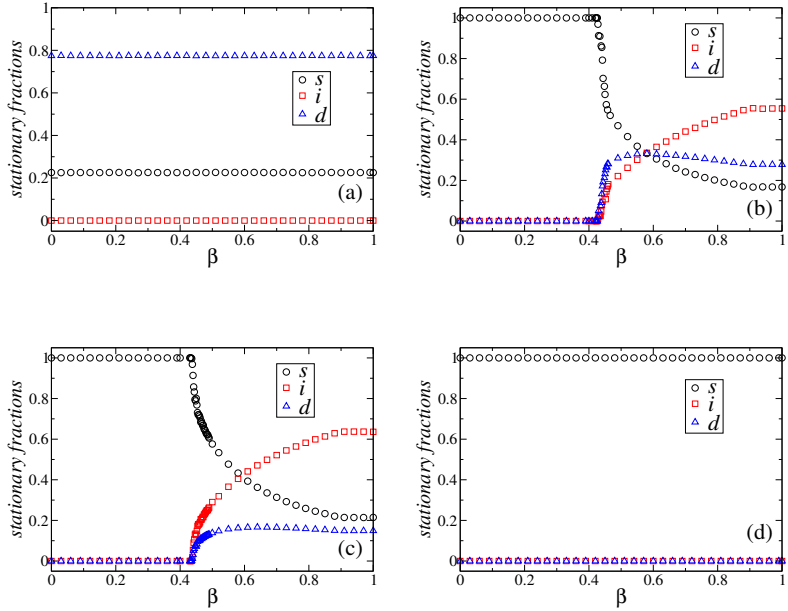

\begin{center}
\vspace{6mm}
\includegraphics[width=0.4\textwidth,angle=0]{figure6a.eps}
\hspace{0.3cm}
\includegraphics[width=0.4\textwidth,angle=0]{figure6b.eps}
\\
\vspace{1.0cm}
\includegraphics[width=0.4\textwidth,angle=0]{figure6c.eps}
\hspace{0.3cm}
\includegraphics[width=0.4\textwidth,angle=0]{figure6d.eps}
\end{center}
\caption{Stationary fractions $s, i$ and $d$ as functions of the contagion probability $\beta$ for the model on the top of BA networks, obtained from agent-based numerical simulations. The fixed parameters are $\alpha=0.10, \delta=0.05$ and $\eta=0.12$, and we varied $\mu$: (a) $\mu=0.02$, (b) $\mu=0.05$, (c) $\mu=0.10$ and (d) $\mu=0.30$. Results are averaged over $100$ independent simulations.} 
\label{fig6}
\end{figure}

The results of the numerical simulations of the model defined for the BA network are shown in Fig. \ref{fig6}, where the stationary fractions $s$, $i$ and $d$ are shown as functions of the contagion probability $\beta$ for different values of the ``refresh rate'' $\mu$. All results shown in this section are averaged over 100 independent realizations, each with randomized initial conditions. The simulations were performed on the same Barab\'asi-Albert network instance, generated as described above, and each time step consists of $N$ update attempts, where $N$ is the number of nodes.

For the lowest value of $\mu=0.02$ studied here, the system evolves to a steady state with only agents in the $S$ and $D$ states, as can be seen in Fig. \ref{fig6}a. This is similar to what was observed for the previous case (see Fig. \ref{fig3}a), although with slightly higher concentration of $S$ agents here. For the intermediate values of $\mu$ (namely $0.05$ and $0.10$, as seen in Figs. \ref{fig6}b and \ref{fig6}c), the qualitative behavior is also similar to what was observed in the fully-connected network: there is a very clear transition, with the increase of the contagion parameter $\beta$, from an absorbing to an active phase. While in Fig. \ref{fig3}b the observed phase constituted of two types of agents, $S$ and $D$, here only $S$ agents are present. Meanwhile, for $\mu=0.10$ (see Fig. \ref{fig3}c) the behavior is quite the same. Of course, the overall proportions of the three states in the active phase are not quantitatively equal, but they are quite similar. If one can point an overall difference present for all cases discusses so far, the deniers population stabilize always in lower values for the BA network. When we increase the ``refresh rate'' $\mu$, though, the BA differs considerably from the mean field - as seen in Fig. \ref{fig6}d, for $\mu=0.30$ the system is in a full absorbing state, with $i,d=0, \forall \beta$. This means that with the increase of the ``refresh rate'' the network resets itself faster than the connectivity between the nodes allows the contagion to spread. On the other hand, $\mu=0.30$ means that 30\% of the population can die and be born/come-of-age at each time step, which may not be too realistic when combined with the slower interactions of the BA network. Therefore, the resetting of $I$ and $D$ agents back to $S$ via the $\mu$ parameter becomes much faster than the propagation conditions and the absorbing $S$ state predominates quickly.

In our simulations, we used the standard BA construction with $m=1$, which yields a minimal scale-free topology. Although we did not vary this parameter systematically, preliminary observations suggest that the qualitative features of the dynamics - such as the presence of an active phase and absorbing states - remain robust. A more detailed investigation of how the model's behavior depends on network construction parameters (e.g., $m$) is left for future work.

It is important to note that the critical values derived in the mean-field formulation are not expected to hold quantitatively for scale-free networks. Nevertheless, our simulations suggest that the same qualitative phases persist under network heterogeneity, including the existence of absorbing and active regimes. A more detailed investigation of critical behavior in scale-free topologies is left for future research.

\subsection{Watts-Strogatz (WS) small-world network}

\quad To assess the robustness of the results under alternative topological conditions, we simulated the model on Watts-Strogatz (WS) networks, which are known to capture the small-world effect, which is a hallmark of many real-world social networks. In particular, we generated WS networks with $N$ nodes following the standard procedure \cite{watts1998collective}. The construction starts with a regular ring lattice where each node is connected to its $k$ nearest neighbors. Then, each edge is randomly rewired with a probability $p$, creating long-range shortcuts that reduce the average path length while preserving a high clustering coefficient. This generates a small-world structure, which better captures features of real social networks. Each independent trial uses the same WS network instance. In the simulations, we considered $k=4$ and $p=0.05$.

The dynamics occurs in the same way described in section 3.2. For the simulations performed here, we considered WS networks with a size of $N=5000$ nodes. At $t=0$, $2\%$ of all agents are in the $I$ state, randomically distributed in the network, while the remaining start at the $S$ state. Although the WS network is naturally more realistic than the fully-connected one, the temporal evolution of the WS network is very similar to the ones shown in Fig. \ref{fig2}, and are therefore not shown here.

\begin{figure}[t]
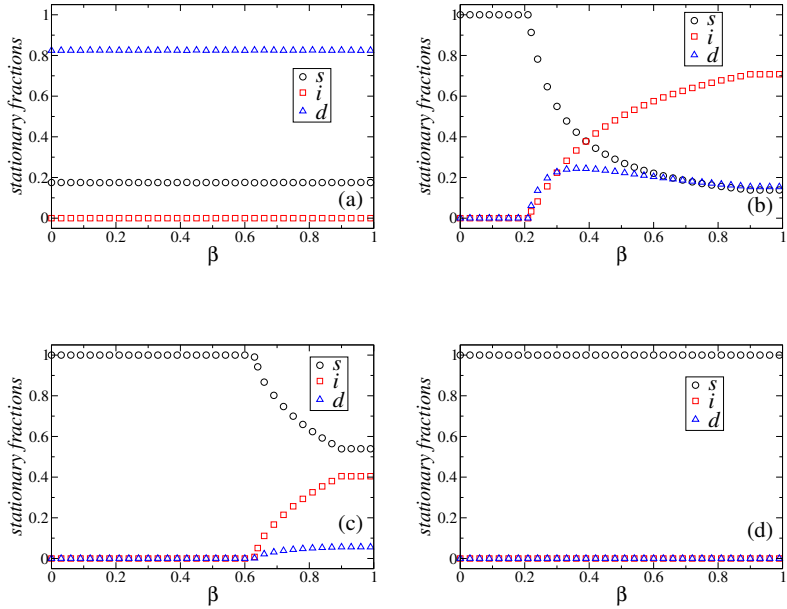

\begin{center}
\vspace{6mm}
\includegraphics[width=0.4\textwidth,angle=0]{figure7a.eps}
\hspace{0.3cm}
\includegraphics[width=0.4\textwidth,angle=0]{figure7b.eps}
\\
\vspace{1.0cm}
\includegraphics[width=0.4\textwidth,angle=0]{figure7c.eps}
\hspace{0.3cm}
\includegraphics[width=0.4\textwidth,angle=0]{figure7d.eps}
\end{center}
\caption{Stationary fractions $s, i$ and $d$ as functions of the contagion probability $\beta$ for the model on the top of WS networks, obtained from agent-based numerical simulations. The fixed parameters are $\alpha=0.10, \delta=0.05$ and $\eta=0.12$, and we varied $\mu$: (a) $\mu=0.02$, (b) $\mu=0.10$, (c) $\mu=0.30$ and (d) $\mu=0.50$. Results are averaged over $100$ independent simulations.} 
\label{fig7}
\end{figure}

The results of the numerical simulations of the model defined for the WS network are shown in Fig. \ref{fig7}, where the stationary fractions $s$, $i$ and $d$ are shown as functions of the contagion probability $\beta$ for different values of the ``refresh rate'' $\mu$, using the same parameter set employed in the Barab\'asi-Albert (BA) simulations. All results shown in this section are averaged over 100 independent realizations, each with randomized initial conditions. The simulations were performed on the same Watts-Strogatz network instance, generated as described above, and each time step consists of $N$ update attempts, where $N$ is the number of nodes.

For the lowest value of $\mu=0.02$ studied here (see Fig. \ref{fig7}a), the dynamics is dominated by the deniers, with no infected nodes and a small fraction of suscetibles. This is similar to what was observed for the previous cases (see Figs. \ref{fig3}a and \ref{fig6}a), although with slightly higher concentration of $S$ agents here in comparison with the mean-field case, and similar to the values observed in the BA network case. The spread of racist content is absent for long times. For the intermediate values of $\mu$ (namely $0.10$ and $0.30$, as seen in Figs. \ref{fig7}b and \ref{fig7}c), the system presents the previously observed continuous phase transitions, and we observe a growing fraction of spreaders of racist content and a minimal presence of deniers. Finally, for $\mu=0.50$, the system reaches the absorbing state where only susceptibles survive in the stationary regime, and the diffusion of racist content is largely contained, as can be seen in Fig. \ref{fig7}d.

These qualitative behaviors closely mirror those observed in the BA network (see Fig. \ref{fig6}), suggesting that the essential features of the model are preserved across different network topologies. Although the critical values of $\beta$ for the transitions slightly shift due to the structural differences between BA and WS networks, the suppression mechanism via deniers remains effective. This reinforces the generality of the model's conclusions.

\section{Final Remarks}

\quad In this work, we proposed and analyzed a mathematical model for the spread of racism in social networks, considering fully-connected, Barabási-Albert (BA) and Watts-Strogatz (WS) networks. Our findings suggest the existence of three macroscopic stationary states: two racism-free phases and one endemic phase, where racist views persist in the system. The transition between these phases depends on network topology and the interaction dynamics among individuals.

The results suggest that the structure of the underlying social network may influence the persistence or eradication of racist ideologies. In particular, the presence of hubs in scale-free networks, as modeled by the Barab\'asi-Albert (BA) topology, indicates that a small number of highly connected individuals can disproportionately affect the system's dynamics, potentially contributing to the emergence of "frozen prejudices" in individuals \cite{galam_trump1,galam_trump2,galam_entropy_2025}. This observation implies that targeted interventions - such as reducing the influence of central nodes or promoting counter-narratives at strategic locations - could be more effective than random interventions. A more systematic investigation of how specific structural properties affect the long-term behavior of the model is left for future work.

We also extended our analysis to Watts-Strogatz (WS) small-world networks, which capture features such as high clustering and short average path lengths, properties commonly observed in empirical social networks. The model exhibits qualitatively similar behavior to that found on Barabási-Albert and fully-connected networks, with the same three macroscopic stationary states and continuous phase transitions. While critical points may shift due to topological differences, the persistence of the general phenomenology across these distinct topologies suggests that the model’s predictions are robust to network structure. This further supports the relevance of the proposed thresholds and intervention strategies across different online environments.

Despite its insights, the model has limitations. The assumption of fixed transition probabilities may oversimplify real-world dynamics, where individual behavior is influenced by external factors such as media, social policies, and offline interactions. Additionally, the model does not explicitly account for feedback loops where institutional or platform-level moderation policies could alter the spread of racist content. Future extensions could incorporate adaptive network structures, stochastic elements in agent behavior, or empirical data calibration to improve the model’s applicability to real-world scenarios. On the other hand, most models found in the literature for racism spreading rely on spontaneous transitions, where in this model we used only social interactions, since interactions predominate in social media, therefore treating racism as a social disease, justifying the SID model. By excluding spontaneous transitions, we also managed to downsize the number of parameters of the system. A further development to be explored is the possibility of policies against racist messages on social media, but since it is something, at least so far, very hard to achieve successfully, we did not include it in the model. Another simplification in our model concerns the direct transition from a susceptible individual to a denier after exposure to racist content (i.e., the $S + I\to D + I$ pathway). This mechanism assumes that some individuals may immediately reject racist ideologies upon first contact. While this allows us to model immediate resistance reactions, it may not capture the more common, gradual processes through which individuals develop critical awareness. Similar constructs have been adopted in models of rumor and information spreading to represent early skepticism or spontaneous resistance to harmful content \cite{ieee,zanette}. Future extensions could incorporate exposure history, cumulative influence, or social reinforcement to better represent these dynamics.

While our study is primarily theoretical, the findings do suggest actionable insights for policy design. For instance, the strong influence of hubs in scale-free networks points to the potential effectiveness of targeted interventions. Moderating or counteracting the influence of highly connected individuals - who may disproportionately amplify racist content - could significantly alter the spread dynamics. This supports the notion that network-aware strategies, rather than uniform ones, might be more effective in combating online racism \cite{rmp_epidemic,prl_cohen}.

Our simulations and analytical results reveal that the transition from low to high prevalence of racist behavior occurs around well-defined critical values of the spreading parameters, notably the infection probability $\beta$ and the reinforcement probability $\eta$. These critical points can serve as thresholds for intervention: if $\beta$ is reduced below the first critical point or if $\eta$ is sufficiently small, the system is driven toward an absorbing state where racism vanishes. Moreover, when $\mu$ is low and $\beta$ lies between the two critical points, the system enters a phase where deniers dominate alongside susceptible individuals, and the racist behavior disappears from the dynamics. This suppression results primarily from the reinforcing interactions among deniers, which prevent the reactivation of infected nodes. While the parameter $\alpha$ does not explicitly enter the critical expressions, it contributes to the proliferation of deniers, enhancing resistance to the spread of hate content. Additionally, since the Barab\'asi-Albert topology reveals that hub nodes act as persistent spreaders, interventions that target high-degree users, for example by moderating their content more strictly or limiting the visibility of their posts, could significantly impact the system’s macroscopic behavior. In contrast, the Watts-Strogatz topology exhibits higher clustering and short average path lengths, features that facilitate local reinforcement among deniers and inhibit long-range propagation of racist content. This suggests that enhancing local counter-speech or peer-based interventions may be particularly effective under such network structures. Thus, our model offers a framework to identify quantitative intervention strategies even in the absence of empirical calibration.

This is reflectd by the consistency of our results across three distinct topologies, mean-field approximation, Barabási-Albert networks and Watts-Strogatz small-world networks, which reinforces the robustness of our model and its relevance to real-world scenarios. In particular, the persistence of absorbing states and critical thresholds suggests that timely intervention strategies can prevent the widespread dissemination of harmful ideologies. From a policy perspective, our findings imply that: (i) Targeted interventions on influential nodes (e.g., users with high connectivity or reach) are likely to be most effective (ii) Community-level moderation or disruption of shortcut pathways, as represented in WS networks, may help suppress undesirable macroscopic outcomes; (iii) Global efforts to reduce susceptibility or enhance denunciation dynamics (e.g., content moderation, reporting tools) can shift the system toward safer states.

Although our study focuses specifically on racism as a form of online social contagion, the broader framework we propose captures key features that underlie the spread of hate speech and extremist ideologies in digital spaces, such as  xenophobia, misogyny or anti-LGBTQ+ rhetoric. In an era where social media platforms often amplify divisive and harmful content, understanding the mechanisms through which such narratives propagate is not merely an academic exercise, it is a societal imperative. Our findings demonstrate how structural properties of networks, such as the presence of influential hubs and small-world effects, can significantly affect the persistence or suppression of toxic discourse. These insights may serve as a theoretical basis for public policy debates, particularly regarding the regulation of digital platforms, algorithmic moderation strategies and the ethical responsibilities of technology companies. In this sense, our model contributes not only to the field of statistical physics but also to the ongoing global conversation on how to balance freedom of expression with the prevention of social harm in increasingly polarized online environments.

Overall, our findings provide a theoretical framework for understanding the dynamics of racism in online networks and emphasize the need for targeted intervention strategies. Further research integrating real-world data and sociopsychological factors could enhance the model’s predictive power and contribute to more effective policies for combating online hate speech.

Although our study primarily investigates the theoretical dynamics of the model and its phase behavior under different parameter regimes, we acknowledge the importance of grounding parameter values in empirical observations. In real-world social media environments, estimating such parameters requires access to detailed interaction data and user behavior logs. Recent works have attempted to bridge this gap by combining agent-based or compartmental models with data-driven techniques to calibrate transition probabilities in social contagion processes, particularly in domains such as online hate speech propagation, misinformation diffusion and digital activism.

For instance, empirical studies often leverage social media platforms' APIs (Application Programming Interfaces) to extract time-stamped interaction patterns, exposure histories, and content dissemination metrics, which can then be used to fit parameters such as $\beta$ (the probability of adopting a behavior after exposure) and $\eta$ (the resistance or denial rate). While such data are typically platform-dependent and subject to privacy constraints, methods like approximate Bayesian computation, likelihood-free inference or machine learning-assisted parameter fitting can be employed to estimate effective rates from observed cascades and user transition trajectories.

Recent empirical analyses of hate‑speech propagation on social media platforms (e.g., \cite{Mathew,Franco,Kentmen-Cin,Pulgarin}) indicate that content tagged as hateful diffuses faster and reaches larger audiences than neutral content, and that the likelihood of a user reacting to or producing hate speech depends heavily on their prior exposure, centrality and interaction history, which are characteristics consistent with the contagion and conversion parameters of the present model. This is why the model focus on propagation in social networks only, despite other sources of exposure and interaction being always present, such as family and school/work environments.

Finally, we acknowledge that the adoption of anti-racist positions is typically the result of long-term educational efforts and sustained social reinforcement. In our model, we opted for a simplified rule in which susceptible individuals can transition to the denier state through a single interaction, following the structure of classical probabilistic contagion models. While this assumption facilitates analytical treatment and aligns with standard approaches in the modeling of opinion and information spreading, we recognize that it may not fully capture the complexity of real-world ideological change. Future extensions could consider cumulative exposure mechanisms, threshold-based dynamics or memory effects to provide a more nuanced representation of anti-racism adoption \cite{dodds,centola}. In addition, we aim in the future to further explore such calibration strategies by integrating empirical data into the model, especially to estimate interaction-driven rates. While our current choice of parameters reflects a stylized exploration of possible system behaviors, the framework is general and adaptable to data-informed studies.


\appendix

\section{Analytical calculations}

Considering the stationary states of the model, we can take the $t\to\infty$ limit in Eqs. (\ref{eq5}) - (\ref{eq7}). From Eq. (\ref{eq6}), we obtain
\begin{eqnarray} \label{eq_a_1}
i & = & 0 \\ \label{eq_a_2}
s & = & \frac{\delta\,d+\mu}{\beta}
\end{eqnarray}
\noindent
From Eq. (\ref{eq5}), we have
\begin{equation} \label{eq_a_3}
s = \frac{\mu}{(\alpha+\beta)\,i + \eta\,d+\mu}
\end{equation}
Finally, from Eq. (\ref{eq7}), we obtain
\begin{equation} \label{eq_a_4}
d=\frac{\alpha\,s\,i}{\mu-\delta\,i-\eta\,s}
\end{equation}

Considering the $i=0$ solution from Eq. (\ref{eq_a_1}), Eq. (\ref{eq_a_4}) gives us: $d=0$. From the normalization condition, $s+i+d=1$, we obtain $s=1$. This first solution, $(s_1,i_1,d_1)=(1,0,0)$ represents the absorbing state of the model.

Considering yet $i=0$ and Eq. (\ref{eq_a_3}), the normalization conditions lead us to a second order polynomial for $d$, namely $\eta\,d^{2}+(\mu-\eta)\,d=0$. This polynomial has two solutions, $d=0$ (which again leads to the absorbing state solution) and $d=1-\mu/\eta$. Thus, the normalization condition gives us the stationary solution for $s$, namely $s=1-\mu/\eta$. Thus, the second solution of the model is given by $(s_2,i_2,d_2)=(1-\mu/\eta,0,\mu/\eta)$.

Finally, the coexistence state where the three compartments survive in the long-time limit can be obtained if we consider a solution $i\neq 0$ in Eqs. (\ref{eq_a_2}) - (\ref{eq_a_4}). Considering Eq. (\ref{eq_a_2}) and the normalization condition, one can obtain
\begin{equation}\label{eq_a_5_new}
d=\frac{\beta-\mu-\beta\,i}{\beta+\delta} = d(i)
\end{equation}

Now, we can plug Eq. (\ref{eq_a_5_new}) in Eq. (\ref{eq_a_2}) to obtain
\begin{equation}\label{eq_a_6_new}
s=\frac{\delta}{\beta}\left(\frac{\beta-\mu}{\beta+\delta}\right) + \mu + \beta\,i = s(i)
\end{equation}

Considering Eqs. (\ref{eq_a_5_new}) and (\ref{eq_a_6_new}) in Eq. (\ref{eq_a_3}), one can obtain the solution $i_3$. Let us call this solution $i_3=i_{3}^{*}$. Thus, the other fractions $d_3$ and $s_3$ can be obtained from Eqs. (\ref{eq_a_5_new}) and (\ref{eq_a_6_new}), respectively. The third model's solution is then given by

\begin{eqnarray}  \label{eq_a_5}
s_3 & = & \frac{\delta}{\beta}\left(\frac{\beta-\mu}{\beta+\delta} + \frac{\mu}{\delta}\right) + \beta\,i_{3}^{*} \\ \label{eq_a_6}
i_3 & = & i_{3}^{*} \\ \label{eq_a_7}
d_3 & = & \frac{\beta-\mu-\beta\,i_{3}^{*}}{\beta+\delta} 
\end{eqnarray}

We can also consider the stability analysis of the above solutions. First, we can write the Jacobian matrix of the model from Eqs. (\ref{eq5}) - (\ref{eq7}),
\begin{equation}
J = 
\begin{pmatrix}
-\beta\,i-\eta\,d-\alpha\,i-\mu & -(\beta+\alpha)\,s & -\eta\,s \\
\beta\,i & \beta\,s-\delta\,d-\mu & \delta\,i \\
\eta\,d +\alpha\,i & \delta\,d+\alpha\,s & \delta\,i+\eta\,s-\mu
\end{pmatrix}
\end{equation}

First we can consider the absorbing-state solution given by $s=1, i=d=0$. Considering this solution, the Jacobian matrix presents the eigenvalues 
$\lambda_1=-\mu, \lambda_2=\beta-\mu$ and $\lambda_3=\eta-\mu$. We have $\lambda_1<0$, and the other two eigenvalues are negative for $\mu>\beta$ and $\mu>\eta$. These are the conditions for the stability of the absorbing solution, and it defines the first critical point, namely
\begin{equation}
\beta_c^{(1)} = \mu
\end{equation}

Considering now the second solution, $s=\mu/\eta, i=0$ and $d=1-\mu/\eta$, two of the eigenvalues are given by the roots of the second-order polynomial $\lambda^{2} + \eta\,\lambda+\mu\,(\eta-\mu)=0$. The Routh-Hurwitz criteria states that a second degree polynomial $P(s)=s^{2}+a_1\,s+a_0$ has roots with negative real parts, and the system $P(s)=0$ is then stable, if and only if both coefficients satisfy $a_i>0$ \cite{rw_criteria}. We have $a_1=\eta>0$ and $a_0=\eta-\mu$ is positive if $\mu<\eta$. This last condition is the opposite we obtained for the absorbing-state solution. 

Yet concerning the second solution, the third eigenvalue is given by 
\begin{equation}
\lambda_3 = \frac{\mu}{\eta}\,(\beta+\delta) - (\delta+\mu)
\end{equation}
\noindent
We have $\lambda_3<0$ for $\beta < (\delta+\mu)\,\frac{\mu}{\eta} - \delta$, which defines the second critical point,
\begin{equation}
\beta_c^{(2)} = (\delta+\mu)\,\frac{\eta}{\mu} - \delta
\end{equation}
Thus, the solution  $s=\mu/\eta, i=0$ and $d=1-\mu/\eta$ is stable for $\beta<\beta_c^{(2)}$, since $\mu<\eta$.

Considering the above two critical points, we observe that the absorbing state solution will be valid for $\mu>\eta$, since $\beta<\beta_c^{(1)}$. On the other hand, the other solution is valid for $\mu<\eta$, since $\beta<\beta_c^{(2)}$. Equating the expressions for the two critical points, they coalesce at $\mu=\eta$. This value separates both solutions with $i=0$. In addition to these two distinct solutions for $i=0$, there is also the above-discussed third solution where $i\neq 0$, valid for $\beta>\beta_c^{(1)}$ and $\beta>\beta_c^{(2)}$.


\section*{Acknowledgments}
N. Crokidakis acknowledges partial financial support from the Brazilian scientific funding agency Conselho Nacional de Desenvolvimento Científico e Tecnológico, Brazil (CNPq, Grant 308643/2023-2).

\vspace{0.5cm}

\noindent
\textbf{Conflict of interest} The authors declare that they have no conflict of interest.

\vspace{0.5cm}

\noindent
\textbf{Data Availability Statement} This manuscript has no associated data or the data will not be deposited.

\vspace{0.5cm}

\noindent
\textbf{Author contribution statement} All authors contributed to the study conception and design. Simulations were performed by N. Crokidakis. Data analysis and interpretation were conducted by all authors. The manuscript was written, reviewed, and approved by all authors.

\bibliographystyle{spphys}       

\end{document}